\documentclass[aps,prd,onecolumn,groupedaddress,showpacs,nofootinbib,amssymb
]{revtex4}
\usepackage[dvips]{graphicx}
\usepackage{amssymb}
\usepackage{amsmath}
\usepackage{graphicx}
\usepackage{amsfonts}
\usepackage{bm}

\begin{document}

\title{A Smooth Constant-Roll to a Slow-Roll Modular Inflation Transition}
\author{V.K. Oikonomou,$^{1,2}$\,\thanks{v.k.oikonomou1979@gmail.com}}
\affiliation{ $^{1)}$ Laboratory for Theoretical Cosmology, Tomsk
State University of Control Systems
and Radioelectronics (TUSUR), 634050 Tomsk, Russia\\
$^{2)}$ Tomsk State Pedagogical University, 634061 Tomsk, Russia\\
}

\begin{abstract}
In this work, we investigate how a smooth transition from a
constant-roll to a slow-roll inflationary era may be realized, in
the context of a canonical scalar field theory. We study in some
detail the dynamical evolution of the cosmological system, and we
investigate whether a stable attractor exists, both numerically and
analytically. We also investigate the slow-roll era and as we
demonstrate, partially compatibility of the resulting scalar theory,
may be achieved, with the potential of the latter belonging to a
class of modular inflationary potentials. The novel features of the
constant-roll to slow-roll transition which we achieved, are firstly
that the it is not compelling for the slow-roll era to last for
$N\sim 50-60$ $e$-foldings, but it may last for a smaller number of
$e$-foldings, since some $e$-foldings may occur during the
constant-roll era. Secondly, when the slow-roll era occurs after the
constant-roll era, the graceful exit from inflation may occur, a
feature absent in the constant-roll scenario, due to the stability
properties of the final attractor in the constant-roll case.
\end{abstract}

\pacs{04.50.Kd, 95.36.+x, 98.80.-k, 98.80.Cq,11.25.-w}

\maketitle



\def\pp{{\, \mid \hskip -1.5mm =}}
\def\cL{\mathcal{L}}
\def\be{\begin{equation}}
\def\ee{\end{equation}}
\def\bea{\begin{eqnarray}}
\def\eea{\end{eqnarray}}
\def\tr{\mathrm{tr}\, }
\def\nn{\nonumber \\}
\def\e{\mathrm{e}}

\section{Introduction}

The inflationary era is unarguably one of the successes of modern
theoretical physics, and in the literature there exist various
approaches, see Refs.
\cite{inflation1,inflation2,inflation3,inflation4,inflation5} for
some reviews, and also Refs.
\cite{Nojiri:2006ri,Capozziello:2011et,Capozziello:2010zz,Nojiri:2010wj,Clifton:2011jh,Nojiri:2017ncd}
for modified gravity descriptions of the inflationary era. The first
approach, and quite common, uses a slow-rolling canonical scalar
field \cite{inflation1,inflation2}, and recently the observational
data coming from Planck \cite{planck}, indicated that plateau-like
potentials are preferred. Although some single scalar field models
of inflation are observationally viable, the single scalar field
description of inflation, leaves no room for non-Gaussianities in
the power spectrum. Up to date, the various distinct modes of the
primordial curvature perturbations are assumed to be uncorrelated,
which yields the Gaussian property of the power spectrum. The
Gaussian property of the primordial modes is up to date supported by
the observational data, however, if non-Gaussianities are detected
in the future, this will put the single scalar field of inflation
models in question. For a recent review on non-Gaussianities, see
Ref. \cite{Chen:2010xka}.

A solution to this issue for single  scalar field models, is to
modify the slow-roll condition, and allow a constant-roll era for
the canonical scalar field
\cite{Inoue:2001zt,Tsamis:2003px,Kinney:2005vj,Namjoo:2012aa,Martin:2012pe,Motohashi:2014ppa,Cai:2016ngx,Hirano:2016gmv,Cook:2015hma,Anguelova:2015dgt,Kumar:2015mfa,Odintsov:2017yud,Odintsov:2017qpp,Nojiri:2017qvx,Fei:2017fub,Gao:2017owg}.
In some previous works \cite{Odintsov:2017yud,Odintsov:2017qpp} we
demonstrated that it is possible to combine a constant-roll with a
slow-roll era, and achieve a smooth transition between these eras.
Particularly in Ref. \cite{Odintsov:2017yud} we showed that it is
possible to achieve a smooth transition between a slow-roll era at
early times, and a constant-roll era at the latest stages of the
inflationary era. Also in Ref. \cite{Odintsov:2017qpp} we
demonstrated that it is possible to achieve a transition between to
constant-roll eras. However, in both cases, the inflationary era
ends up to a constant-roll cosmological attractor, so it was not
possible to end this inflationary era, since the attractor was
stable. To this end, in this paper we shall demonstrate how to
achieve a transition between a constant-roll era, which occurs at
the early stages of the inflationary era, and a slow-roll era, which
occurs at the late stages of the inflationary era. The interesting
features of this approach is that firstly the slow-roll era is not
required to last for $N\sim 50-60$ $e$-foldings, but possibly lasts
for a smaller number of $e$-foldings, since some $e$-foldings may
occur during the constant-roll era. Secondly, with the slow-roll era
occurring after the slow-roll era, one may obtain an inflationary
model which is compatible with the Planck \cite{planck} data, but
also the graceful exit from inflation is possible. As we will
demonstrate, the model we shall present can have compatibility with
the Planck data, however we could not achieve simultaneous
compatibility of both the spectral index and the scalar-to-tensor
ratio. Interestingly enough, the resulting slow-roll model has a
potential that belongs to a class of modified modular inflation
\cite{BenDayan:2008dv}.

This paper is organized as follows: In section II, we shall briefly
present the essential features of the transition mechanism between
constant and slow-roll eras, and also we discuss several issues
regarding the cosmological dynamics. In section III, we shall
present the model of transition from a constant to slow-roll era,
and we shall study the dynamical properties of the cosmological
system. We investigate if the dynamical solution we obtain can drive
the transition ,and we validate if this solution is the final
attractor, both numerically and analytically. Finally, we calculate
the observational indices and we examine the parameter space in
order to see when compatibility with the observational data may be
achieved. Finally, the conclusions follow in the end of the paper.

Before we start, we need to clarify the choice of the geometric
background which we will adopt in the rest of this paper. The
geometric background is assumed to be a flat
Friedmann-Robertson-Walker metric with line element, \be
\label{metricfrw} ds^2 = - dt^2 + a(t)^2 \sum_{i=1,2,3}
\left(dx^i\right)^2\, , \ee with $a(t)$ being the scale factor.
Also, the connection is assumed to be a metric compatible, symmetric
and torsion-less affine connection, the Levi-Civita connection.

\section{Essential Features of Dynamical Transitions Between Constant-Roll Inflation and Slow-Roll Inflation}

In order to maintain the article self-contained, we shall briefly
present the formalism we developed in our previous works
\cite{Odintsov:2017yud,Odintsov:2017qpp}. We shall consider a
canonical scalar field theory, with the geometric background being a
flat FRW one, with the scalar action being,
 \be
\label{canonicalscalarfieldaction}
\mathcal{S}=\sqrt{-g}\left(\frac{R}{2}-\frac{1}{2}\partial_{\mu}\phi
\partial^{\mu}\phi -V(\varphi))\right)\, ,
\ee where $V(\varphi)$ is the canonical scalar potential. The
corresponding scalar energy density is equal to, \be
\label{energydensitysinglescalar}
\rho=\frac{1}{2}\dot{\varphi}^2+V(\varphi)\, , \ee and in addition,
the corresponding pressure is,
 \be \label{pressuresinglescalar}
P=\frac{1}{2}\dot{\varphi}^2-V(\varphi)\, . \ee In effect, the
Friedmann equation is, \be \label{friedmaneqnsinglescalar}
H^2=\frac{1}{3 M_p^2}\rho\, , \ee and it is easy to show that,
 \be
\label{dothsinglescalar} \dot{H}=-\frac{1}{2M_p^2}\dot{\varphi}^2\,
. \ee Moreover, the canonical scalar field obeys the Klein-Gordon
equation,
 \be \label{kleingordonsingle}
\ddot{\varphi}+3H\dot{\varphi}+V'=0\, , \ee where the prime denotes
differentiation with respect to $\varphi$.

In most inflationary theories, the inflationary dynamical evolution
is determined by the slow-roll parameters, and for the scalar theory
these are $\epsilon$ and $\eta$, which are actually the leading
order terms in the perturbative expansion known as Hubble slow-roll
expansion \cite{Liddle:1994dx}. The slow-roll parameters $\epsilon$
and $\eta$ are defined as follows,
 \be
\label{slowrollindiceshubblerate} \epsilon=-\frac{\dot{H}}{H^2}\,
,\quad \eta=-\frac{\ddot{H}}{2H\dot{H}}\, , \ee and also these can
be written in terms of the canonical scalar field as follows,
 \be
\label{slowrollindiceshubblerate123}
\epsilon=\frac{\dot{\varphi}^2}{2M_p^2H^2}\, ,\quad
\eta=-\frac{\ddot{\varphi}}{2H\dot{\varphi}}\, . \ee Recently, the
slow-roll theoretical framework was extended, and in the new
framework known as constant-roll inflation
\cite{Martin:2012pe,Motohashi:2014ppa}, the second slow-roll index
$\eta$, may become a constant number $\eta=-n$, which is not
required to be small in magnitude \cite{Martin:2012pe}. The
constant-roll inflationary framework was further extended in Refs.
\cite{Odintsov:2017yud,Odintsov:2017qpp}, in which, transition
between constant and slow-roll eras are possible. This framework is
based on the basic assumption that the second slow-roll index $\eta$
has the following form,
\begin{equation}\label{basciccnd1}
\eta=-f(\varphi (t))\, ,
\end{equation}
which can be written as follows,
\begin{equation}\label{basiccondition}
\frac{\ddot{\varphi}}{2H\dot{\varphi}}=f(\varphi (t))\, ,
\end{equation}
where function $f(\varphi (t))$ in Eqs. (\ref{basciccnd1}) and
(\ref{basiccondition}) is considered to be a smooth and monotonic
function of $\varphi (t)$, and also it has to be dimensionless. Also
the functional behavior will determine the transition between
constant and slow roll eras.

As in our previous works we shall use the Hamilton-Jacobi formalism,
and our aim is to find the Hubble rate expressed in terms of the
canonical scalar $H(\varphi )$, by solving the corresponding
equations of motion. An important aim is to check whether the
solution $H(\varphi )$ is the final attractor of the cosmological
dynamical system, regardless of the choice of the initial
conditions, and we will check this both analytically and numerically
in the next section. Let us present in brief some of the basic
equations of the cosmological dynamical system we shall use. Due to
the fact that,
\begin{equation}\label{sense}
\dot{H}=\dot{\varphi}\frac{\mathrm{d}H}{\mathrm{d}\varphi}\, ,
\end{equation}
Eq. (\ref{dothsinglescalar}) can be cast as follows,
\begin{equation}\label{extra1}
\dot{\varphi}=-2M_p^2\frac{\mathrm{d}H}{\mathrm{d}\varphi}\, ,
\end{equation}
and by combining Eqs. (\ref{extra1}) and (\ref{basiccondition}),
after some algebra we obtain the following,
\begin{equation}\label{masterequation}
\frac{\mathrm{d}^2H}{\mathrm{d}\varphi^2}=-\frac{1}{2M_p^2}f(\varphi
)H(\varphi )\, .
\end{equation}
The differential equation (\ref{masterequation}) is of central
importance in this paper, since this will determine the function
$H(\varphi)$, which may or may not be the final attractor of the
cosmological dynamical system. Having the function $H(\varphi)$ at
hand, the scalar potential can easily be found by combining Eqs.
(\ref{extra1}) and (\ref{friedmaneqnsinglescalar}), and it reads,
\begin{equation}\label{potentialhubblscalar}
V(\varphi)=3M_p^2H(\varphi)^2-2M_p^4(H'(\varphi))^2\, .
\end{equation}
In the next section, we shall use the formalism we developed in this
paper, and as we will show, we shall achieve a smooth transition
from a constant to a slow-roll era.

\section{A Smooth Transition from a Constant-roll era to a Slow-roll Era}

The choice of the function $f(\varphi (t))$ plays a crucial role in
order for the transitions to occur, and it is conceivable to notice
that during the constant-roll era, the function $f(\varphi (t)$ must
be a constant, and during the slow-roll era, it has to be
approximately $f(\varphi (t)) \simeq 0$. In order to achieve a
transition from a constant-roll era to a slow-roll era, we shall
assume that the function  $f(\varphi (t))$ appearing in Eq.
(\ref{basiccondition}), has the following form,
\begin{equation}\label{choice1}
f(\varphi (t))=\frac{\alpha  \sinh (\lambda x)}{\gamma+\alpha \sinh
(\lambda  x)}\, ,
\end{equation}
where for the purposes of this work $\lambda$ is assumed to be equal
to $\lambda=\frac{\sqrt{\frac{1}{2}}}{M_p}$, and the parameters
$\gamma$ and $\alpha$ are dimensionless parameters. The choice of
$f(\varphi (t)$ appearing in Eq. (\ref{choice1}) dictates that the
constant-roll era occurs for large field values, when
$\frac{\varphi}{M_p}\gg 1$, and as the field values decrease, when
$\varphi<M_p$ the slow-roll era commences. Hence, the slow-roll era
which we will discuss in this paper, belongs to the small field
inflationary models. For large field values, the function $f(\varphi
(t))$ behaves as, $f(\varphi \gg M_p)\sim 1$, while for small field
values, it behaves as $f(\varphi \ll M_p)\sim 0$, so for small field
values a slow-roll era is realized. Hence, the constant-roll era
corresponds to the scenario studied in Ref.
\cite{Motohashi:2014ppa}, with  $\alpha=-4$, if we use the notation
of Ref. \cite{Motohashi:2014ppa}.

\begin{figure}[h]
\centering
\includegraphics[width=20pc]{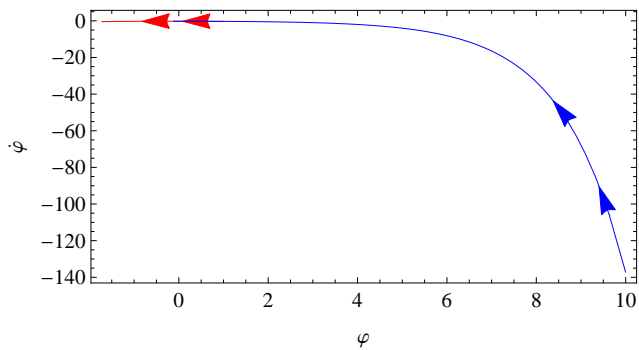}
\caption{The phase space diagram of the solution
(\ref{generalsolution1}), for the initial conditions $\varphi
(0)=10^{20}$ (blue curve) and $\varphi (0)=10^{19}$ (red
curve).}\label{plotsnumerics1}
\end{figure}

Let us now proceed by finding the solution of the differential
equation (\ref{masterequation}), with the function $f(\varphi (t))$
being chosen as in Eq. (\ref{choice1}), and the resulting solution
for $H(\varphi )$ is,
\begin{equation}\label{generalsolution1}
H(\varphi)=\gamma +\alpha  \sinh \left(\frac{\varphi }{2
M_p^2}\right)\, .
\end{equation}
Accordingly, the scalar potential $V(\varphi)$ can be found, by
combining Eqs. (\ref{generalsolution1}) and
(\ref{potentialhubblscalar}), and it is equal to,
\begin{equation}\label{potentialcase1}
V(\varphi)= M_p^2 \left(-2 \alpha ^2+3 \gamma ^2+\alpha ^2 \cosh
\left(\frac{\sqrt{2} \varphi }{M_p}\right)+6 \alpha \gamma \sinh
\left(\frac{\varphi }{\sqrt{2} M_p}\right)\right)\, .
\end{equation}
The potential can be further simplified for small field values, so
during the slow-roll era of the scalar field $\varphi$, and the
simplified potential reads,
\begin{equation}\label{approx1potential}
V(\varphi)\simeq 3 \gamma ^2 M_p^2-\alpha ^2 M_p^2 \cosh
^2\left(\frac{\varphi }{\sqrt{2} M_p}\right)\, ,
\end{equation}
and by further expanding the second term in powers of the scalar
field, in the small $\varphi/M_p$ limit, we obtain,
\begin{equation}\label{potentialapproxfinal}
V(\varphi)\simeq -\frac{\alpha ^2 \varphi ^2}{2}+3 \gamma^2
M_p^2-\frac{\alpha ^2 \varphi ^4}{12 M_p^2}-\alpha ^2 M_p^2\, .
\end{equation}
The potential (\ref{potentialapproxfinal}) is a type of modular
inflationary potential, which can be found in Ref.
\cite{BenDayan:2008dv}, and the general form of the potential is,
\begin{equation}\label{generalform1}
V(\varphi)=\alpha_1-\alpha_2 \varphi^2-\alpha_3\varphi^q\, .
\end{equation}
Obviously, the potential (\ref{potentialapproxfinal}) corresponds to
the case $q=4$ of the potential (\ref{generalform1}). As we show
later on this section, the potential (\ref{potentialapproxfinal})
leads to an interesting phenomenology, partially compatible with the
2015 Planck data \cite{planck}. Before getting to that, we need to
investigate whether the solution $H(\varphi)$ is the final attractor
of the cosmological dynamical system, so we will check this both
numerically and analytically. We start off with the numerical
approach, so we choose the following values for the parameters
$\alpha$ and $\gamma$ (the reasons for this choice will become clear
later on),
\begin{equation}\label{choicesofparameters}
\alpha=0.165,\,\,\,\gamma=2\, ,
\end{equation}
and also recall that $\lambda=\frac{\sqrt{\frac{1}{2}}}{M_p}$. For
the values of the parameters chosen as in
(\ref{choicesofparameters}), in Fig. \ref{plotsnumerics1} we present
the phase space diagram $\dot{\varphi}-\varphi$ corresponding to the
solution (\ref{generalsolution1}), by using the initial conditions
$\varphi (0)=10^{20}$ (blue curve) and $\varphi (0)=10^{19}$ (red
curve).
It is obvious from Fig. (\ref{plotsnumerics1}), that the solution
(\ref{generalsolution1}) is the final attractor of the cosmological
dynamical system, since an attractor behavior occurs, regardless the
choice of initial conditions. We need to note that the final
attractor has appealing physical properties, since at the field
values decrease, the velocity of the field $\dot{\varphi}$ also
decreases.

\begin{figure}[h]
\centering
\includegraphics[width=20pc]{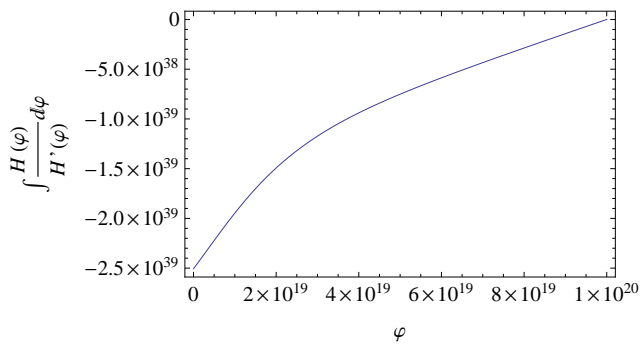}
\caption{Stability of linear perturbations for the solution
$H(\varphi)=\gamma +\alpha  \sinh \left(\frac{\varphi }{2
M_p^2}\right)$.}\label{stabilityperturbations}
\end{figure}

Apart from the above numerical study for the stability of the
attractor, we can also investigate it's stability analytically. We
shall adopt the technique developed in Ref. \cite{Liddle:1994dx}, so
we vary Eq. (\ref{potentialhubblscalar}), and we get the following
differential equation with respect to $\delta H(\varphi)$,
\begin{equation}\label{perturbationeqnsbbasic1}
H_0'(\varphi)\delta H'(\varphi)\simeq
\frac{3}{2M_p^2}H_0(\varphi)\delta H(\varphi)\, ,
\end{equation}
where $H_0(varphi)$ is the solution (\ref{generalsolution1}). The
general solution of the differential equation
(\ref{perturbationeqnsbbasic1}) is,
\begin{equation}\label{perturbationsolution1}
\delta H(\varphi )=\delta
H(\varphi_0)e^{\frac{3}{2M_p^2}\int_{\varphi_0}^{\varphi}\frac{H_0(\varphi)}{H_0(\varphi
)}}\, ,
\end{equation}
where the parameter $\varphi_0$ denotes some initial value of
$\varphi$. Then, having the solution $H_0(\varphi)$ at hand, we can
investigate whether this solution is a stable final attractor of the
cosmological system, by studying the evolution of the linear
perturbations (\ref{perturbationsolution1}) as a function of
$\varphi$. Clearly, if the perturbations grow, the solution
$H_0(\varphi)$ s unstable, and on the contrary case, it proves to be
stable, at list at a linear perturbation theory context. The sign of
the exponential term in (\ref{perturbationsolution1}), determines
whether the perturbations grow (positive sign), or decay (negative
sign). In Fig. \ref{stabilityperturbations} we present the behavior
of the terms appearing in the exponential terms, as a function of
$\varphi$.
As it can be seen, the term in the exponent is negative, hence the
perturbations decay, and therefore the solution
(\ref{generalsolution1}) is stable towards linear perturbations, and
thus it is the final attractor of the theory.

Having discussed the attractor properties of the solution
(\ref{generalsolution1}), let us now focus on the slow-roll era, and
we examine the viability of the canonical scalar field model
(\ref{potentialapproxfinal}), by confronting the corresponding
observational indices with the 2015 Planck data \cite{planck}. The
potential in the limit $\varphi<M_p$ is given in Eq.
(\ref{potentialapproxfinal}), so by using this we will calculate the
slow-roll indices $\epsilon$ and $\eta$, which in terms of the
potential in the context of the slow-roll approximation, can be
written as follows,
\begin{equation}\label{slowrollscalar}
\epsilon (\varphi)=\frac{M_p^2}{2}\left(
\frac{V'(\varphi)}{V(\varphi)}\right)^2,\,\,\,\eta
(\varphi)=M_p^2\frac{V''(\varphi)}{V(\varphi)}\, .
\end{equation}
Having  $\epsilon$ and $\eta$ at hand, we can calculate the
corresponding observational indices, namely the spectral index of
the primordial curvature perturbations and the scalar-to-tensor
ratio. The latter two in the case of a canonical scalar field, can
be written in the following form,
\begin{equation}\label{spectrali}
n_s\simeq 1-6\epsilon+2\eta,\,\,\,r=16 \epsilon\, .
\end{equation}
By using the definition of the $e$-foldings number,
\begin{equation}\label{efoldings}
N\simeq \frac{1}{M_p^2}
\int_{\varphi_f}^{\varphi_{i}}\frac{V(\varphi)}{V'(\varphi)}\mathrm{d}\varphi
\, ,
\end{equation}
where $\varphi_i$ is the value of the scalar field at the horizon
crossing, and also by determining the final value of the scalar
field when $\epsilon\sim 1$, we can obtain, after some extensive
algebra, the spectral index $n_s$, which at leading order is,
\begin{align}\label{nsfinalform}
& n_s\simeq \frac{2 \alpha ^2 M_p^2 \left(\left(-\frac{6 \gamma
^2}{\alpha ^2}+\frac{\sqrt{3} \mathcal{Q}(\alpha,\gamma)}{\alpha ^4
M_p^2}+1\right) e^{\frac{2 \alpha ^2 N}{\alpha ^2-3 \gamma
^2}}-1\right)}{\alpha ^2 \left(-M_p^2\right)+3 \gamma ^2
M_p^2+\frac{\left(\sqrt{3} \mathcal{Q}(\alpha,\gamma)+M_p^2
\left(\alpha ^4-6 \alpha ^2 \gamma ^2\right)\right) e^{\frac{2
\alpha ^2 N}{\alpha ^2-3 \gamma ^2}}}{2 \alpha
^2}-\frac{\left(\sqrt{3} \mathcal{Q}(\alpha,\gamma)+M_p^2
\left(\alpha ^4-6 \alpha ^2 \gamma ^2\right)\right)^2 e^{\frac{4
\alpha ^2 N}{\alpha ^2-3 \gamma ^2}}}{12 \alpha ^6 M_p^2}}+1\\
\notag & \frac{48 \alpha ^4 M_p^2 \left(M_p^2 \left(\frac{6 \gamma
^2}{\alpha ^2}-1\right)-\frac{\sqrt{3}
\mathcal{Q}(\alpha,\gamma)}{\alpha ^4}\right) e^{\frac{2 \alpha ^2
N}{\alpha ^2-3 \gamma ^2}} \left(3 M_p^2+\left(M_p^2 \left(\frac{6
\gamma ^2}{\alpha ^2}-1\right)-\frac{\sqrt{3}
\mathcal{Q}(\alpha,\gamma)}{\alpha ^4}\right) e^{\frac{2 \alpha ^2
N}{\alpha ^2-3 \gamma ^2}}\right)^2}{\left(\alpha ^2 \left(12
M_p^4+6 M_p^2 \left(M_p^2 \left(\frac{6 \gamma ^2}{\alpha
^2}-1\right)-\frac{\sqrt{3} \mathcal{Q}(\alpha,\gamma)}{\alpha
^4}\right) e^{\frac{2 \alpha ^2 N}{\alpha ^2-3 \gamma
^2}}+\frac{\left(\sqrt{3} \mathcal{Q}(\alpha,\gamma)+M_p^2
\left(\alpha ^4-6 \alpha ^2 \gamma ^2\right)\right)^2 e^{\frac{4
\alpha ^2 N}{\alpha ^2-3 \gamma ^2}}}{\alpha ^8}\right)-36 \gamma ^2
M_p^4\right)^2}
\end{align}
where $\mathcal{Q}(\alpha,\gamma)$ stands for,
\begin{equation}\label{fgdpotra}
\mathcal{Q}(\alpha,\gamma)=\sqrt{\alpha ^6 \left(-M_p^4\right)
\left(\alpha ^2-4 \gamma ^2\right)}\, .
\end{equation}
Accordingly, the scalar-to-tensor ration can be found, and it's
analytic expression at leading order is,
\begin{align}\label{scalarrrrr}
& r\simeq \frac{128 \alpha ^4 M_p^2 \left(M_p^2 \left(\frac{6 \gamma
^2}{\alpha ^2}-1\right)-\frac{\sqrt{3}
\mathcal{Q}(\alpha,\gamma)}{\alpha ^4}\right) e^{\frac{2 \alpha ^2
N}{\alpha ^2-3 \gamma ^2}} \left(3 M_p^2+\left(M_p^2 \left(\frac{6
\gamma ^2}{\alpha ^2}-1\right)-\frac{\sqrt{3}
\mathcal{Q}(\alpha,\gamma)}{\alpha ^4}\right) e^{\frac{2 \alpha ^2
N}{\alpha ^2-3 \gamma ^2}}\right)^2}{\left(\alpha ^2 \left(12
M_p^4+6 M_p^2 \left(M_p^2 \left(\frac{6 \gamma ^2}{\alpha
^2}-1\right)-\frac{\sqrt{3} \mathcal{Q}(\alpha,\gamma)}{\alpha
^4}\right) e^{\frac{2 \alpha ^2 N}{\alpha ^2-3 \gamma
^2}}+\frac{\left(\sqrt{3} \mathcal{Q}(\alpha,\gamma)+M_p^2
\left(\alpha ^4-6 \alpha ^2 \gamma ^2\right)\right)^2 e^{\frac{4
\alpha ^2 N}{\alpha ^2-3 \gamma ^2}}}{\alpha ^8}\right)-36 \gamma ^2
M_p^4\right)^2}
\end{align}

\begin{figure}[h]
\centering
\includegraphics[width=20pc]{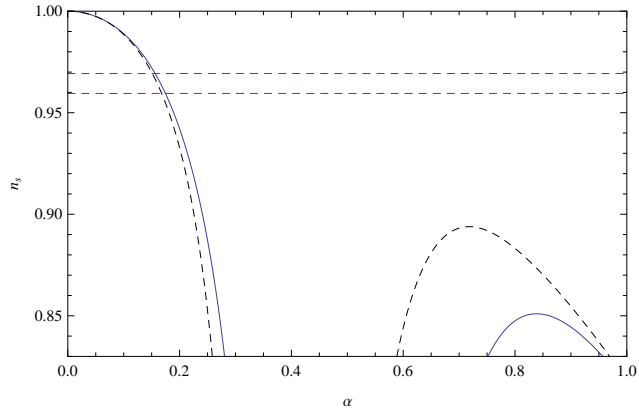}
\caption{The spectral index as a function of $\alpha$ for
$\gamma=2$, and for $N=60$ (black and dashed curve) and $N=40$ (blue
curve). The upper red dashed line is the upper limit of the Planck
allowed value for the spectral index, $n_s=0.9693$ and the lower red
and dashed line, the lower Planck compatible limit on $n_s$, namely
$n_s=0.9595$.}\label{plot3}
\end{figure}
\begin{figure}[h]
\centering
\includegraphics[width=20pc]{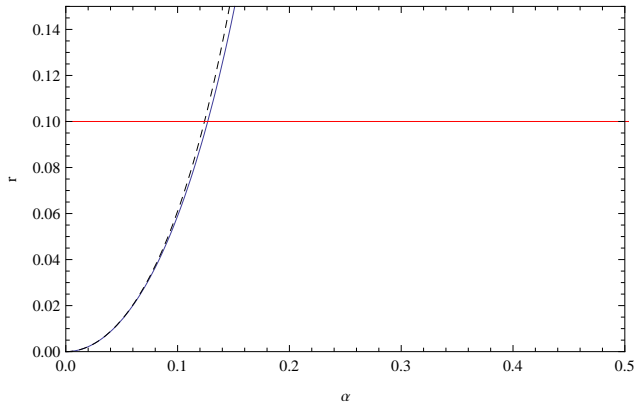}
\caption{The scalar-to-tensor ratio as a function of $\alpha$ for
$\gamma=2$, and for $N=60$ (black and dashed curve) and $N=40$ (blue
curve). The upper red dashed line is the upper limit of the Planck
allowed value for the scalar-to-tensor ratio $r=0.10$}\label{plot4}
\end{figure}

The 2015 Planck data \cite{planck} constrain the observational
indices as follows,
\begin{equation}
\label{planckdata} n_s=0.9644\pm 0.0049\, , \quad r<0.10\, ,
\end{equation}
so now we will investigate the parameter space to see if the
compatibility with the Planck data can be achieved. After thoroughly
analyzing the parameter space, the resulting picture is that
compatibility with the Planck data can be obtained for a large range
of the parameters separately for the spectral index and the
scalar-to-tensor ratio, but not together. For example by choosing
$\gamma=2$ and $\alpha=0.165$, we get $n_s=0.964898$ and
$r=0.185485$, so the scalar-to-tensor ratio is excluded. This
picture seems to persist regardless on the magnitude of the free
parameters. In Fig. \ref{plot3}, we plot the spectral index as a
function of $\alpha$ for $\gamma=2$, and for $N=60$ (black and
dashed curve) and $N=40$ (blue curve). The upper red dashed line is
the upper limit of the Planck allowed value for the spectral index,
$n_s=0.9693$ and the lower red and dashed line, the lower Planck
compatible limit on $n_s$, namely $n_s=0.9595$.
As it can be seen in Fig \ref{plot3}, the compatibility with the
Planck data can be achieved even for $N<50$. This is the new feature
of the present paper, since it is possible to obtain a viable
spectral index inflationary theory, with $N<50$ during the slow-roll
era, since some $e$-foldings may occur during the constant-roll era.
The drawback is that the spectral index and the scalar-to-tensor
ratio are not simultaneously compatible with the Planck data. For
completeness, in Fig. \ref{plot4}, we plot the behavior of the
scalar-to-tensor ratio as a function of the parameter $\alpha$, for
$\gamma=2$, and for $N=60$ (black and dashed curve) and $N=40$ (blue
curve). The red line denotes the upper Planck limit $r=0.10$.
Before closing, we need to stress the two important features of the
present paper. Firstly, we showed that the slow-roll era may last
for a shorter time, so for $N<50$, if some $e$-foldings occur during
the constant-roll era. Secondly, in our approach, a graceful exit is
possible, since it may occur during the slow-roll era, when
$\epsilon, \eta \sim 1$. This was not possible in the constant-roll
framework, since in most cases the constant-roll era is very stable.

\section{Conclusions}

In this work we studied smooth transitions from a constant-roll era
to a slow-roll era. We examined the dynamics of the cosmological
solution, and we investigated whether this solution is a stable and
final attractor of the cosmological dynamical system. We performed
both a numerical and analytical analysis, and as we showed, the
cosmological solution $H(\varphi )$ is a stable attractor and also
it is the final attractor of the system. So the cosmological system
is led to a slow-roll era from a constant-roll era. Also, with
regard to the slow-roll era, we investigated the resulting limiting
form of the potential, which turned out to be a particular form of
modular inflation. We made a thorough analysis of the free
parameters space of the model, and we investigated that the model
can be partially compatible with the Planck data, meaning that we
did not achieve both the spectral index and the scalar-to-tensor
ration to be compatible with the Planck data. However, this is very
much model dependent. The new features that this work suggests are,
firstly, the fact that the slow-roll era comes after the
constant-roll, so in principle the inflationary era may end, since
if the constant-roll era occurs last, it is very stable, so the
possibility of exit is small. Secondly, and more importantly, the
central idea of the paper is that a constant-roll and slow-roll era
may occur together, and it is then possible that some $e$-foldings
of the Universe occur during the constant-roll era, and in effect,
the slow-roll era may last for $N<50-60$ $e$-foldings. Now the
ultimate quest is to find a model which is completely viable and
compatible with the Planck data. We need to note that we examined
various cases which we did not include in the paper, and potentials
that involve exponentials seem to occur if the slow-roll era occurs
before the constant-roll era. We hope to address these issues in a
future paper, and work is in progress towards this direction.

\section*{Acknowledgments}

This work is supported by the Ministry of Education and Science of
Russia Project No. 3.1386.2017 (V.K.O).

\end{document}